\documentclass[epj,nopacs]{svjour}
\usepackage{amsmath,amssymb,graphicx,bm}
\usepackage{color}
\usepackage{multicol}        
\usepackage[bottom]{footmisc}

\newcommand{\beq}{\begin{equation}}
\newcommand{\eeq}{\end{equation}}
\newcommand{\bea}{\begin{eqnarray}}
\newcommand{\eea}{\end{eqnarray}}
\newcommand{\ba}{\begin{align}}
\newcommand{\ea}{\end{align}}
\newcommand{\bfig}{\begin{figure}}
\newcommand{\efig}{\end{figure}}

\newcommand{\D}{\displaystyle}

\newcommand{\rkpisq}{\langle r^2 \rangle_S^{K\pi}} 

\begin{document}


%
\title{Improved bounds on the radius and curvature of the $K\pi$
scalar form factor and implications to low-energy theorems}
\author{Gauhar Abbas \and B.\ Ananthanarayan}

\institute{Centre for High Energy Physics,
Indian Institute of Science, Bangalore\ 560 012, India.\\
{\email{gabbas@cts.iisc.ernet.in} \\
 \email{anant@cts.iisc.ernet.in} \\ 
}}        

\date{April 26, 2009}
%
\abstract{We obtain stringent bounds in the
$\langle r^2\rangle_S^{K\pi}$-$c$ plane where
these are the scalar radius and
the curvature parameters of the scalar $\pi K$ form factor
respectively  using  analyticity and
dispersion relation constraints, 
the knowledge of the form factor from the
well-known Callan-Treiman point $m_K^2-m_\pi^2$, as well as at 
$m_\pi^2-m_K^2$ which we call the second Callan-Treiman point.
The central values of these parameters
from a recent determination are accomodated in the
allowed region provided the higher loop corrections to the
value of the form factor at the second Callan-Treiman point
reduce the one-loop result by about 3\% with  
$F_K/F_\pi=1.21$.  Such a variation in magnitude at
the second Callan-Treiman point  yields
$0.12\, {\rm fm}^2 \lesssim \langle r^2 \rangle_S^{K\pi}
\lesssim 0.21\, {\rm fm}^2$ and $0.56 \, {\rm GeV}^{-4} \lesssim c 
\lesssim 1.47\, {\rm GeV}^{-4}$ 
and a strong correlation
between them.  A smaller value of $F_K/F_\pi$ shifts both bounds 
to lower values.
}
\titlerunning{Improved bounds on the radius and curvature...} 
\authorrunning{Gauhar Abbas \and B. Ananthanarayan}
\maketitle

The scalar $\pi K$ form factor $f_0(t)$, where $t$ is the square
of the momentum transfer, 
is of fundamental importance in
semi-leptonic decays of the kaon and has been studied in great
detail, see, e. g.~\cite{review} for a recent review.  In chiral
perturbation theory it was computed to one-loop accuracy in
ref.~\cite{GL1} and to two-loop accuracy in ref.~\cite{PS,BT}.
It has a branch cut starting at the threshold
$t_+=(m_K+m_\pi)^2$ and is analytic elsewhere in the complex-plane. 
The scalar radius $\langle r^2\rangle^{K\pi}_S$ and the
curvature parameter $c$ arise in the expansion:
\beq\label{expansion}
	f_0(t) = f_+(0)\left(1 +\frac{1}{6} 
\rkpisq
t +
c t^2+ \ldots \right)
\eeq
where $f_+(t)$ is the corresponding vector form factor which
here we normalize to 0.976 as in a recent important work~\cite{JPO} 
that we use for comparison with our results.
An important result on this form factor concerns its value at
the unphysical point $(m_K^2-m_\pi^2)$, known
as the Callan-Treiman (CT) point~\cite{Callan:1966hu} (see also
ref.~\cite{Dashen:1969bh})    
and here it equals
$F_K/F_\pi\simeq 1.21$\footnote{Although now a little
too large, we mainly 
adopt this value in order to compare our results 
with prior results.}, the ratio of the kaon and pion decay constants 
resulting from a soft-pion theorem. It receives very small corrections
at one-loop order which are expected to stay small at higher orders
as well (for recent discussions see ref.~\cite{BG,KN}; note
that in this work we are in the isospin conserving limit).  
These involve coupling constants 
$C_{12}^r$ and $C_{34}^r$ for which there are 
estimates in the literature~\cite{BT,hep-ph/0503108}, and whose
consequences have been discussed at length 
in a recent paper~\cite{arXiv:0903.1654}.  
A soft-Kaon analogue fixes its value at tree-level
at $m_\pi^2-m_K^2$ (which we will refer to as
the second CT point) to be $F_\pi/F_K$~\cite{Oehme}.  
The one-loop correction, $\tilde{\Delta}_{CT}^{NLO}$,
increases the value by $0.03$~\cite{GL1}. 
The rather small size of this correction  may be traced to
the fact that it is parameter free at this level.  The corresponding
correction at two-loop level has been estimated, which gives
the estimate $-0.035 < \tilde{\Delta}_{CT}^{NNLO}
< 0.11$~\cite{arXiv:0903.1654}.  One of the important findings
in our work is that this correction can actually be estimated
using analyticity methods and substantially restricts the
range above, while remaining consistent with it.

Bourrely and Caprini (BC)~\cite{BC} consider
certain dispersion relations for observables 
denoted by $\Psi''(Q^2)$ and $(\Psi(Q^2)/Q^2)'+
\Psi(0)/Q^4$ (which we will name
${\cal O}_1$ and ${\cal O}_2$ respectively)  
involving the square of the form factor.
Employing the
information at the first CT point and phase of the form factor 
along the cut they obtained 
bounds on the scalar radius and curvature parameters. 
(For an accessible introduction to the methods involved see
ref.~\cite{Bourrely:1980gp}.)  Our work inspired by BC,
will use the information at both the CT points to constrain
the expansion coefficients using
the same observables, but will not include the phase information. 
We will find that in order
to accomodate well-known determinations of the same coefficients~\cite{JPO},
the value of the scalar form factor at the second CT point would
have to be lowered by about 3\% compared to its one-loop
value.  This is consistent with
the estimate given in ref.~\cite{arXiv:0903.1654}, and significantly
pins down the correction.  In addition, 
we consider the observable $\Pi'(Q^2)$ 
studied by Caprini~\cite{Caprini} which we denote by ${\cal O}_3$.

To begin, one introduces a conformal variable $z$, 
\beq
z(t)=
S\left(\frac{\sqrt{t_+}-\sqrt{t_+-t}}{\sqrt{t_+}+\sqrt{t_+-t}}\right)
\label{ztmap.eqn1}
\eeq 
where $S$ can be $\pm 1$ depending on the convention (the convention of
BC corresponds to $S=+ 1$, while that of ref.~\cite{Caprini}
corresponds to $S=-1$). 
The relevant dispersion relation is brought into a canonical form: 
\begin{equation}
\frac{1}{2 \pi} \int_{-\pi}^{\pi} d\theta |h(\exp(i\theta))|^2 
\leq I
\end{equation}
where $I$ is the bound and is associated with the observable in question.
In the above, we have
\beq
 h(z) = f(z) w(z),
\label{hz.eqn1}
\eeq
where $f(z)$ is the form factor in terms of the conformal variable
and $w(z)$ is the outer function associated with the
relevant dispersion relation and the Jacobian of the
transformation from $t$ to $z(\equiv \exp(i\theta))$.
The function $h(z)$ then admits an expansion given by
\begin{equation}
h(z)=a_0+a_1 z+ a_2 z^2 + \ldots
\end{equation}
where the $a_i$ are real.
From the Parseval theorem of Fourier analysis, 
we have $a_0^2+a_1^2+a_2^2+...\leq I$.
Improvements on the bound result from additional information which
may be at values of
(1) space-like momenta, or  (2) time-like momenta below threshold where
the form factor is real, or (3) at time-like momenta above threshold
where one may have knowledge either of the modulus or the phase
or both.  Alternatively, if $I$ is known, and the
series is truncated, then one may obtain bounds on the
allowed values of the expansion coefficients of the form factor.  
In BC, the value of the form factor at
the first CT point which belongs to the category (2) above, and the phase
of the form factor between threshold and $1\, {\rm GeV}^2$ in the
region of the type (3) above have been used.  
The improvements result when one takes the constraints one at
a time, and further when they are simultaneously imposed.  
In BC the significant constraint is that from the CT point, in relation to
the constraint from the phase of form factor.  
In the present work the result of wiring in the second
CT constraint alone, and simultaneously with the first one are studied.
We will consider the three observables 
${\cal O}_1$, ${\cal O}_2$ and
${\cal O}_3$.
Their corresponding outer functions are listed in the appendix.
Next we need to consider the following expansion coefficients: 
\beq
a_0 = h(0) = f_+(0) w(0),
\eeq
\beq
a_1 = h^{\prime}(0)=
f_+(0) \left( w^{\prime}(0) +S \D\frac{2}{3} \rkpisq t_+ w(0) 
\right),
\eeq
\bea
a_2 & = &\D\frac{h^{\prime \prime}(0)}{2!}=\frac{f_+(0)}{2}\left[
	w(0)\left(-\frac{8}{3} \rkpisq t_+ + 32 \,c \, t_+^2\right) \right] \nonumber \\ 
	 &+& \frac{f_+(0)}{2} \left[2 w^{\prime}(0)\left(S\frac{2}{3} \rkpisq
 t_+\right) +w^{\prime \prime}(0) \right]. 
\eea
Improving the bounds 
on expansion coefficients subject to constraints from the space-like
region
has been studied recently in the context of the pion electromagnetic
form factor~\cite{AR1}.  The results there are also 
applicable to the case at hand:
our bounds are obtained by solving the determinantal
equation for an observable which in general reads:
\beq
\left|
\begin{array}{c c c c c c }
I & a_0 & a_1 & a_2 &  h(x_1) & h(x_2) \\
a_0 & 1 & 0 & 0  & 1 & 1 \\
a_1 & 0 & 1 & 0  & x_1 & x_2  \\
a_2 & 0 & 0 & 1  & x_1^2 & x_2^2 \\
h(x_1) & 1 & x_1 & x_1^2 &  (1-x_1^2)^{-1} & (1-x_1 x_2)^{-1} \\
h(x_2) & 1 & x_2 & x_2^2 & (1-x_2 x_1)^{-1} & (1-x_2^2)^{-1} \\
\end{array}\right|=0,
\label{det.eqn1}
\eeq
where $x_1$ and $x_2$ are the values of $z$ corresponding to $t=m_K^2-m_\pi^2$
and $t=m_\pi^2-m_K^2$ respectively, and have the numerical values
$x_1= S\times 0.202 , \, x_2=S \times(-0.111)$.  
In the above, observable by observable, we input values for the
quantity $I$.  
Discarding both the rows
and columns corresponding to $x_1$ and $x_2$ would give the bounds
with no constraints, discarding the row and column corresponding
to $x_1$ would amount to including the constraint from only $x_2$ and
{\it vice versa}.  
The results of our analysis are displayed in Figs. 1-4 and in
the discussion below.  For these purposes the value of the
form factor at the first CT point is always taken to be
the ratio $F_K/F_\pi$, while at the second CT point to be
at its one-loop value of $F_\pi/F_K+0.03$, unless otherwise
mentioned.  Unless otherwise mentioned $F_K/F_\pi$ is taken 
to be 1.21 in order to carry out a meaningful comparison
with the results of BC.

In Fig. 1, we display the result obtained when the observable
${\cal O}_1$ is used.  We use as an input for $I_1$ the number 
0.000079 as in ref.~\cite{BC},   
obtained from perturbative QCD
with the choice $Q^2=4 {\rm GeV}^2$, and choice of masses
as given in ref.~\cite{MILC}.  
The result of including the constraint from the second CT point
is truly dramatic isolating a significantly different region  
(the major axes of the two large ellipses are essentially orthogonal). 
This feature is special to this system where one constraint
comes from the time-like yet unphysical region (first CT point)
while the other from a genuine space-like region (second CT point).
Taking the contraints one at a time leads to a small region of intersection
and the ellipse obtained with simultaneous inclusion is even smaller.  
For this case we find the range to be
$0.15 \, {\rm fm}^{ 2} \lesssim \rkpisq \lesssim 0.19 \, {\rm  fm}^{ 2}$
and $0.65 \, {\rm GeV}^{-4}\lesssim c \lesssim 1.35 \, {\rm GeV}^{-4}$,
and we have the approximate relation $c\simeq 19.4 \rkpisq -2.2$ which
is the equation of the major-axis of the ellipse, where $\rkpisq$ is in
fm$^2$ and $c$ is in GeV$^{-4}$. 
As such, it would therefore imply that the curvature effects are not
negligible and must be included in fits to experimental data, as already
observed in, e.g., ref.~\cite{BC,review}.  We have checked that the
ellipse has a non-trivial intersection with the band determined by
the dispersive representation relation between the slope and 
curvature parameters given in eq.~(2.11) of ref.~\cite{review}.
\begin{figure}
\begin{center}
\includegraphics*[angle = 0, width = 0.44\textwidth, clip = true] 
{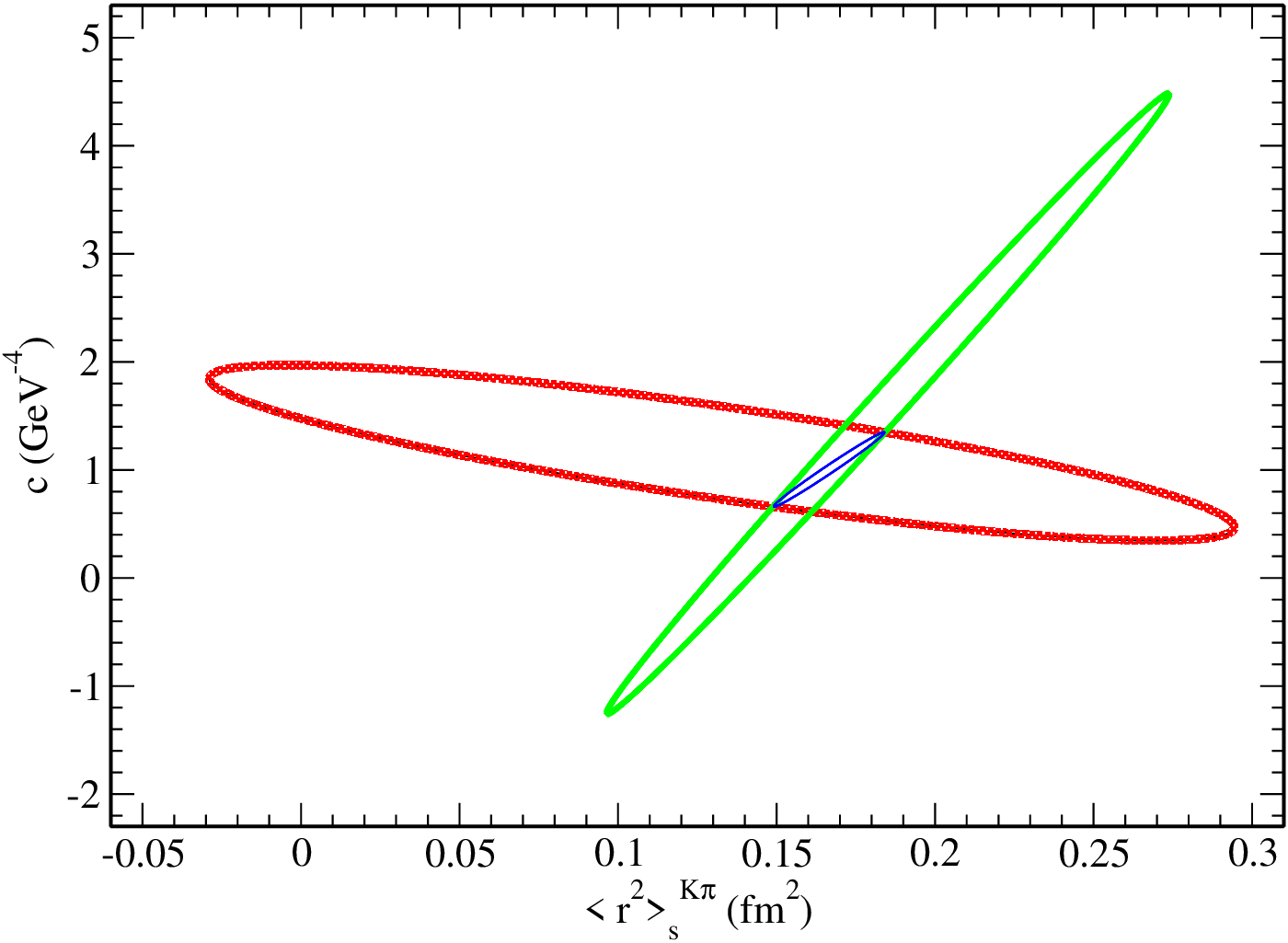}
\caption{Boundaries of the allowed regions in the 
$\langle r^2\rangle_S^{K\pi}$-$c$ plane 
The flat ellipse comes from the constraint at the
first CT point (see also~\cite{BC}), the long narrow ellipse
from the constraint at the second CT point. The small ellipse
results from simultaneous inclusion of both constraints.
}
\end{center}
\end{figure}

The system is sensitive to the value of the form factors
at the CT points.  Since its value at the first CT point is
expected to be stable, we hold it fixed and consider the
variation at the second CT point only.  Although not entirely
consistent as the corrections at both are correlated in chiral
perturbation theory, this is done for purposes of illustration. 
In Fig. 2, we display the effect of varying the value of the
scalar form factor at the second CT point in a 3\% range compared
to its one-loop value.  
Also shown in this figure are the results
of a recent evaluation of the two quantities of interest~\cite{JPO} in the
form a diamond and a cross, following the discussion of BC.  
As the one-loop value is lowered
by 3\%, these are essentially accomodated in the ellipse, which
we consider to be remarkable.  
\begin{figure}
\begin{center}
\includegraphics*[angle = 0, width = 0.45\textwidth, clip = true] 
{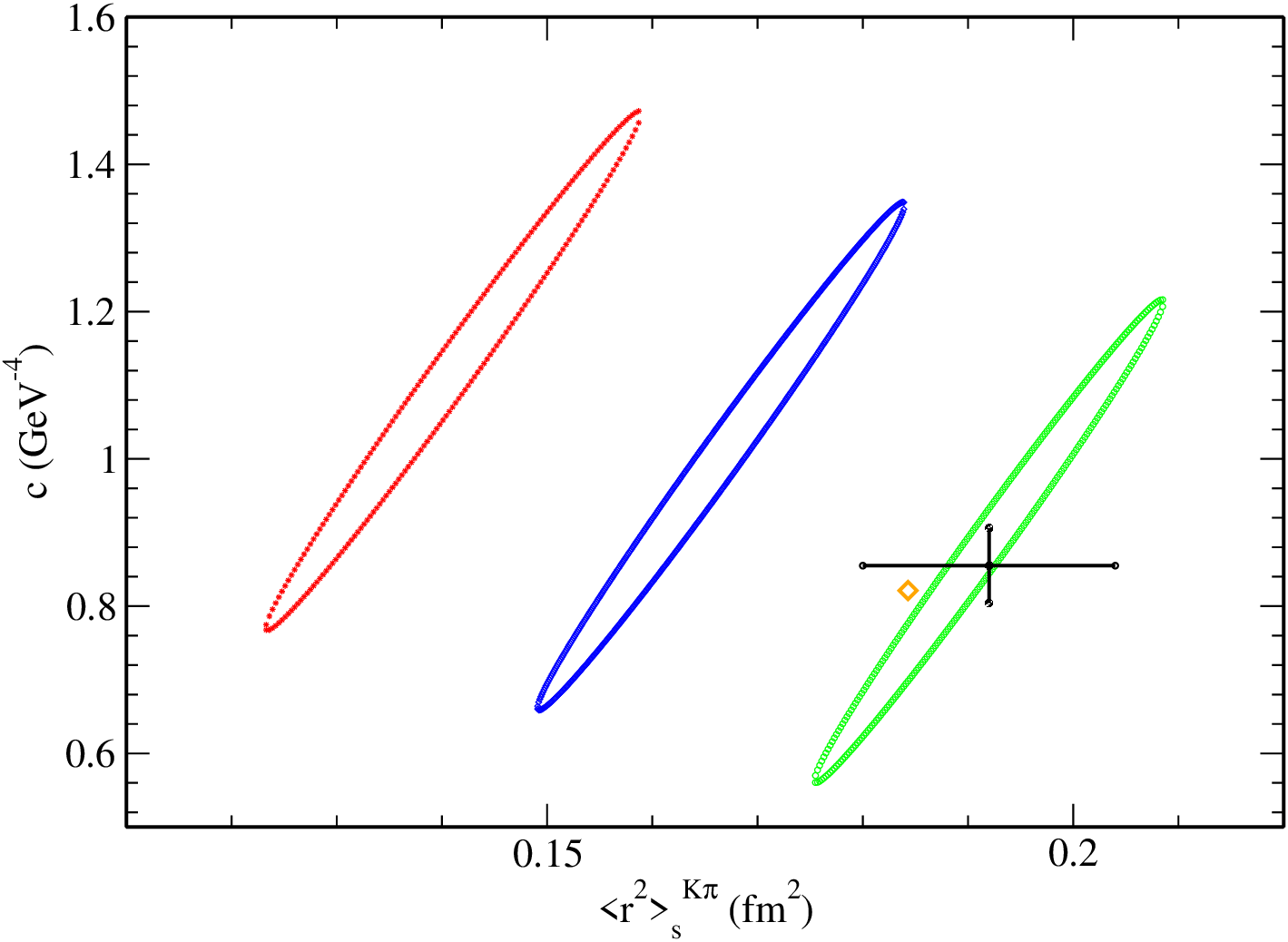}
\caption{Allowed ellipses when the value of the scalar form factor
at the second CT point is changed from the one-loop
result.  The higher ellipse is for
an increase in its value by 3\%, the central ellipse is for when it
is not changed, and the lower ellipse is for when it is lowered by 3\%.
Also marked are the best
values from ref.~\cite{JPO} following \cite{BC} which
essentially lie in the lowest ellipse.}
\end{center}
\end{figure}

We display in Fig. 3 the results obtained by changing (1) the input to
the observable,  and (2) the value
of $F_K/F_\pi$ which is taken to be 1.21 as before, and 1.19, 
as a test of sensitivity to the inputs.  
For the former, we follow 
BC~\cite{BC}: we consider changing the value of the input $I_1$ to the value
0.00020 corresponding to the choice of quark masses from 
ref.~\cite{hep-ph/0409312}.  This choice continues to
be reasonable as a recent determination of quark masses~\cite{Blossier:2007vv} 
yields quark masses numbers
that lie between those of the two prior determinations cited above. 
It may be observed that for $F_K/F_\pi$ of 1.21, even the larger
ellipse does not accomodate the diamond and cross.
We now have
$0.14 \, {\rm fm}^{2} \lesssim \rkpisq \lesssim 0.20 \, {\rm fm}^{2}$
and $0.43 \, {\rm GeV}^{-4}\lesssim c \lesssim 1.58 \, {\rm GeV}^{-4}$. 
\begin{figure}
\begin{center}
\includegraphics*[angle = 0, width = 0.45\textwidth, clip = true] 
{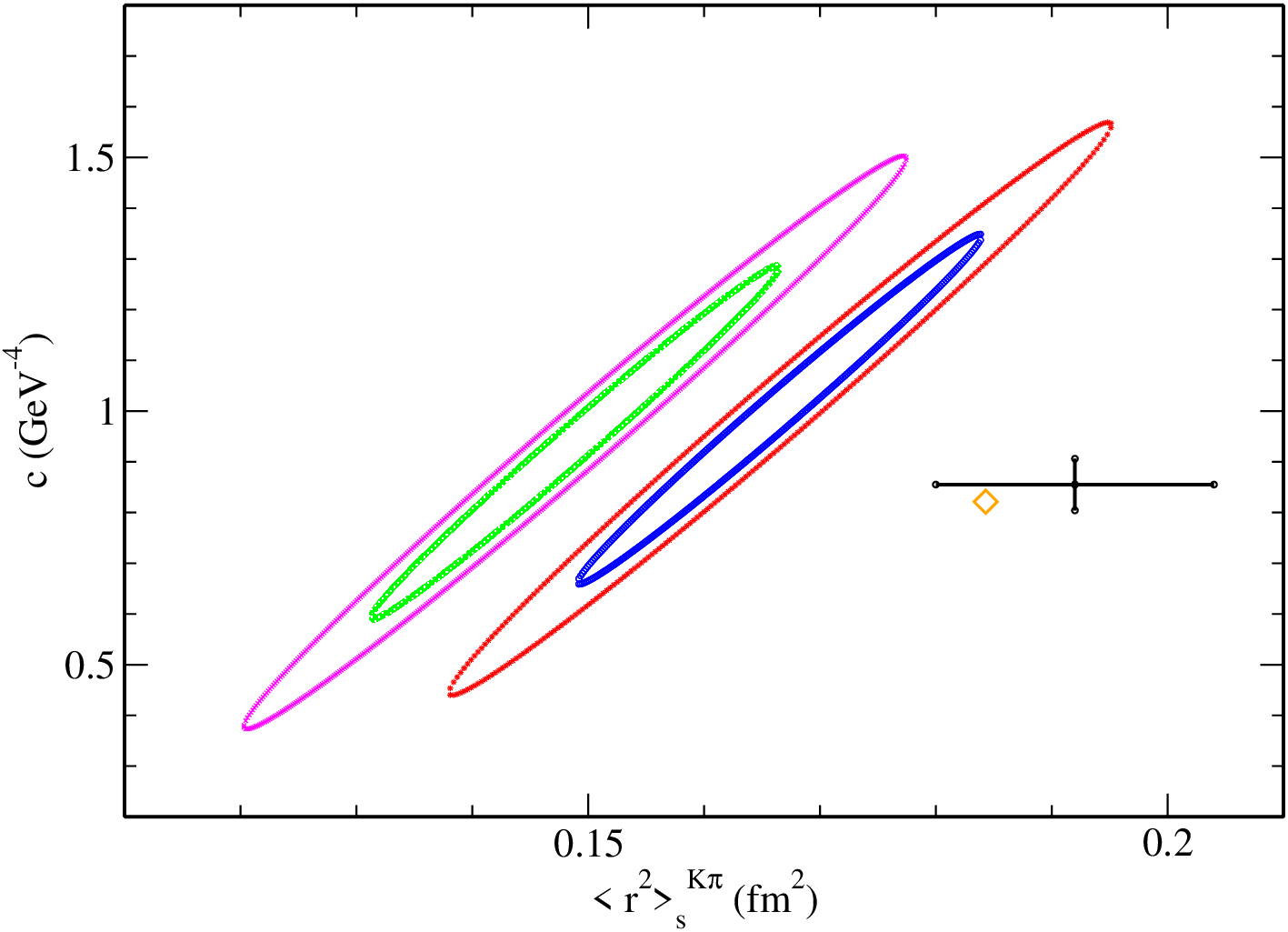}
\caption{
The ellipses for choices for different input values of
$I_1$.  The inner ellipse is for the choice
from ref.~\cite{MILC}, while the outer ellipse for the choice from
ref.~\cite{hep-ph/0409312}, and for choices of $F_K/F_\pi$ of 1.19
and 1.21.  The set of ellipses to the left correspond to the former
while those to the right to the latter. 
Also marked are the best
values from ref.~\cite{JPO} following \cite{BC}.}
\end{center}
\end{figure}

We test our constraints by changing the observables:
as in BC, the first and
second observables are both evaluated with the MILC data~\cite{MILC},   
and we take for the input of the second observable,
$I_2$ the value 0.00022, evaluated at $Q^2=4 {\rm GeV}^2$.
The observable ${\cal O}_3$ has to be adapted to the
problem at hand:
following Caprini, ref.~\cite{Caprini}
we take $I_3$ to be 0.0133 GeV$^{-2}$ $/(m_K^2-m_\pi^2)^2$ 
with $Q^2=2 {\rm GeV}^2$.  
${\cal O}_3$ provides
a much larger allowed region as it is not optimal for the
problem at hand;  the original observable brings in the vector
form factor as well. 
The observables
${\cal O}_1$ and ${\cal O}_2$ essentially isolate the same region
and there is no special advantage in selecting one over the other.
\begin{figure}
\begin{center}
\includegraphics*[angle = 0, width = 0.45\textwidth, clip = true] 
{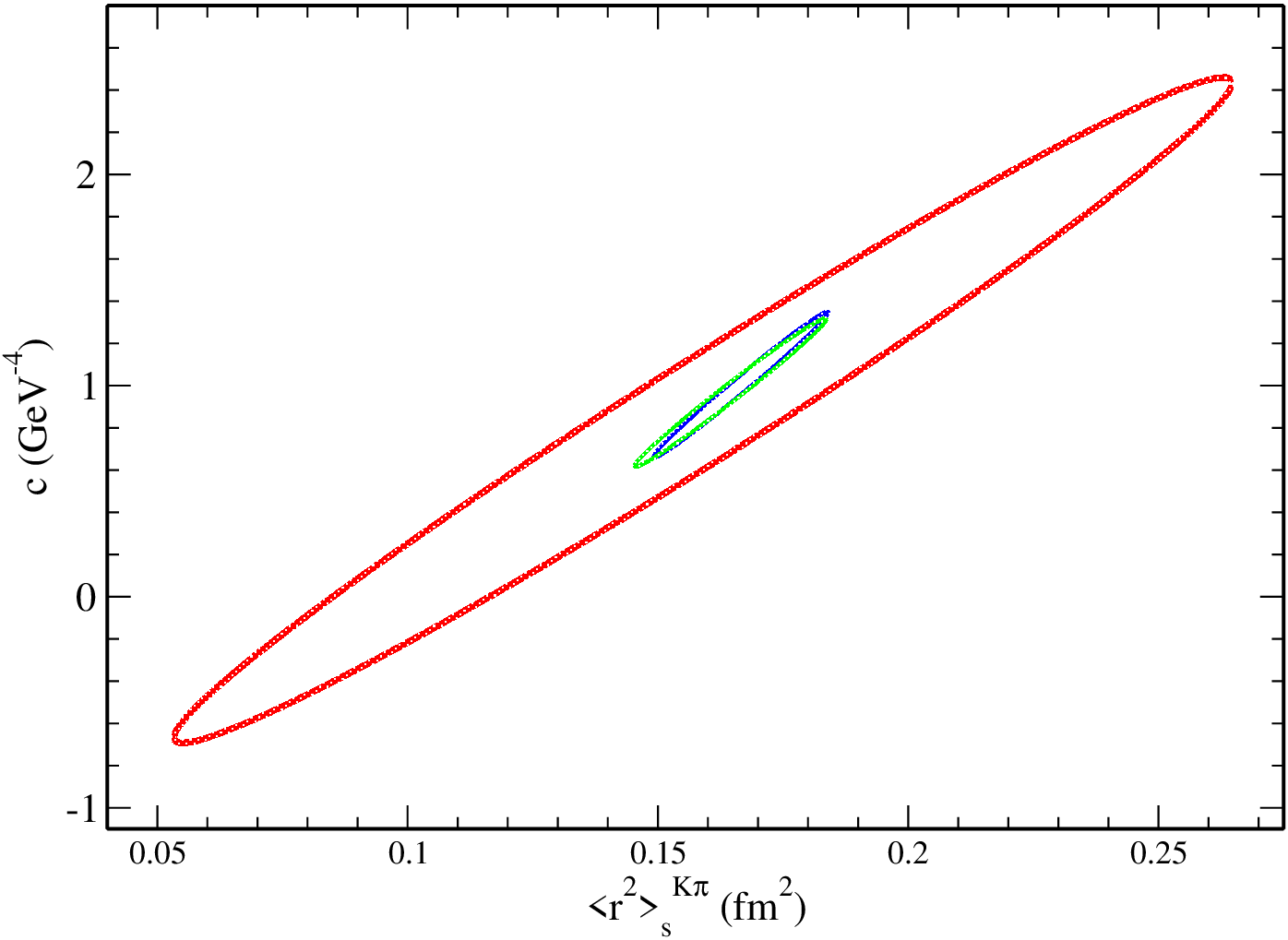}
\includegraphics*[angle = 0, width = 0.45\textwidth, clip = true] 
{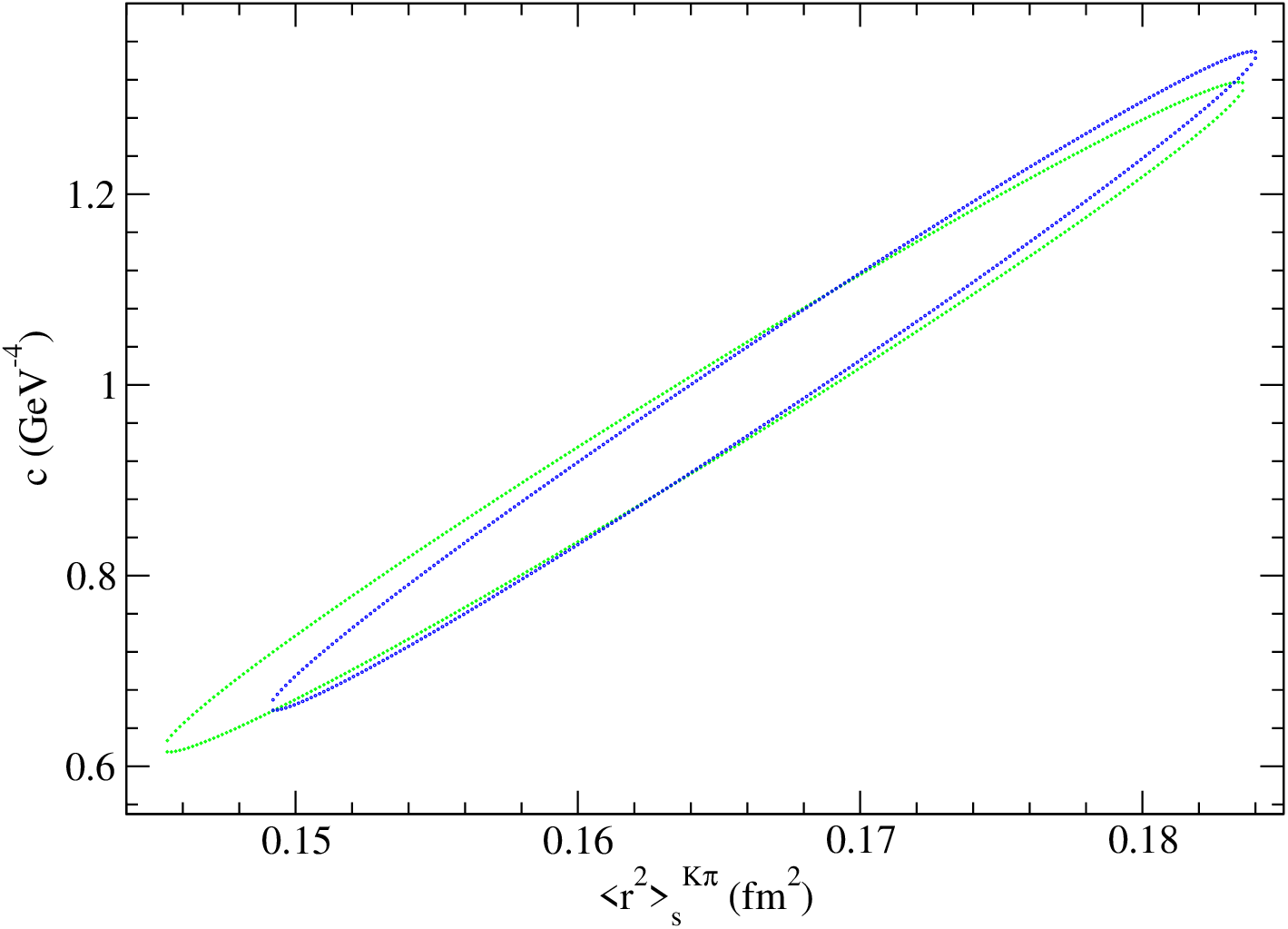}
\caption{Allowed regions shown for different choices of input observables.
In the top panel,
the large ellipse is the allowed region from the observable ${\cal O}_3$.
Of the two smaller ones,
the ellipse that is higher at the right extremity 
is for the observable ${\cal O}_1$, while the other is
for ${\cal O}_2$ (the lower panel zooms into the region of
the latter two ellipses).  }
\end{center}
\end{figure}
In light of the investiations above, the constraints from the
two CT points give the following ranges for the scalar radius
and the curvature parameters: if the ratio of $F_K/F_\pi$ is
fixed to be 1.21, then varying the value of the scalar form factor
at the second CT point by about 3\% from its one-loop value
leads to the ranges
$0.12\, {\rm fm}^2 \lesssim \langle r^2 \rangle_S^{K\pi}
\lesssim 0.21\, {\rm fm}^2$ and $0.56 \, {\rm GeV}^{-4} \lesssim c 
\lesssim 1.47 \, {\rm GeV}^{-4}$ and it may be observed that
there is a strong correlation between the two given by our
ellipses.  On the other hand, if the ratio of the decay constants
is somewhat lower, then the ellipses migrate to the left.
Note that this determination of the radius gives for the
slope parameter $\lambda_0
(\equiv \rkpisq m_\pi^2/6)$ 
the range $10\times 10^{-3}\lesssim \lambda_0
\lesssim 17\times 10^{-3}$ (for a discussion of present day experimental
status see ref.~\cite{review}).
Finally the following may be noted:
the inclusion of phase of
the form factor with (i) the datum
from the second CT point, or (ii) extending the
framework of BC further to include the data from both CT points are
worth studying.  
The analysis also may shed light on issues considered in
many recent studies
e.g., ref.~\cite{KN,Passemaretal}. 
Indeed, BC have constrained the
constants $C_{12}^r$ and $C_{34}^r$;  
our results could also be extended to meet such an end. 
While our work points to a correction of about $-3\%$
to the one-loop value of the scalar form factor due to
higher order effects at the second CT point, 
an interesting analysis would be one that
parallels, e.g., ref.~\cite{JPO} using more current
values of $F_K/F_\pi$.

\begin{acknowledgement}
BA thanks DST for support.  
We are indebted to I. Caprini and H. Leutwyler
for detailed comments on the manuscript,
J. Bijnens, B. Moussallam and E. Passemar for correspondence 
and S. Ramanan for discussions.
\end{acknowledgement}

\section*{Appendix}
In this appendix, we list the relevant outer functions: 
\begin{eqnarray*}
& \displaystyle
w_{{\cal O}_1}=\frac{1}{4} \sqrt{\frac{3}{2 \pi}}\times  & \\
& \displaystyle
\frac{m_K-m_\pi}{m_K+m_\pi}
\frac{(1-z)(1+z)^{3/2} 
\sqrt{1-z+\beta(1+z)}}{(1-z+\beta_Q (1+z))^3}, & \\
& \displaystyle
w_{{\cal O}_2}= w_{{\cal O}_1}\times 
\frac{1-z+\beta_Q(1+z)}{\sqrt{8}}   & \\
\end{eqnarray*}
and
\begin{eqnarray*}
& \displaystyle
w_{{\cal O}_3}=
\frac{(1-d)^2}{32 t_+ (1-z_-)^{5/2}} \sqrt{\frac{3}{4 \pi
t_+}}\frac{(1+z)(1-z)^{5/2}}{(1-z z_-)^{1/4} (1-z d)^2}, &  
\end{eqnarray*}
with
\begin{eqnarray*}
& \displaystyle
\beta=  \sqrt{1-t_-/t_+}, \,
t_-=(m_K-m_\pi)^2,
\beta_Q=  \sqrt{1+Q^2/t_+},   & \\
& \displaystyle 
d = (\sqrt{t_++Q^2}-\sqrt{t_+})/ (\sqrt{t_++Q^2}+\sqrt{t_+}) \, {\rm and}& \\
& \displaystyle 
 z_- = 
(\sqrt{t_+-t_-}-\sqrt{t_+})/ (\sqrt{t_+-t_-}+\sqrt{t_+}). & \\ 
\end{eqnarray*}

\end{document}